\documentclass[prb,twocolumn]{revtex4}

\usepackage{graphicx}

\begin{document}

\title{Ferromagnetic/DMS hybrid structures: one- and zero-dimensional magnetic traps for quasiparticles}

\author{P. Redli\'{n}ski}
\email{Pawel.Redlinski.1@nd.edu} \affiliation{Department of Physics,
University of Notre Dame, Notre Dame, IN 46556}

\author{T. Wojtowicz}
\affiliation{Department of Physics, University of Notre Dame, Notre
Dame, IN 46556} \affiliation{Institute of Physics, Polish Academy of
Sciences, Warsaw, Poland}

\author{T. G. Rappoport}
\affiliation{Department of Physics, University of Notre Dame, Notre
Dame, IN 46556}

\author{A. Libal}
\affiliation{Department of Physics, University of Notre Dame, Notre
Dame, IN 46556}

\author{J. K. Furdyna}
\affiliation{Department of Physics, University of Notre Dame, Notre
Dame, IN 46556}

\author{B. Jank\'{o}}
\affiliation{Department of Physics, University of Notre Dame, Notre
Dame, IN 46556}

\begin{abstract}
We investigated possibility of using local magnetic field
originating from ferromagnetic island deposited on the top of
semiconductor quantum well to produce zero- and one-dimensional
traps for quasi-particles with spin. In particular we considered two
shapes of experimentally made magnets - cylindrical and rectangular.
In the case of ferromagnetic micro-disk the trap can localize spin
in three dimensions, contrary to the rectangular micro-magnet which
creates a trap that allows free propagation in one direction. We
present in detail prediction for absorption spectrum around the main
absorption edge in both type of micro-magnets.
\end{abstract}

\maketitle



Recently there is an increasing interest in using the spin of
particles, instead of their charge, as a basis for the operation of
a new type of electronic devices. In this work we show via
theoretical calculations that spin degrees of freedom can be
utilized for achieving spatial localization of both charged
quasi-particles (electrons, holes, trions) as well as of neutral
complexes (excitons). Such localized states are of interest from
spintronic application point of view.

The hybrid structure we consider is build of \mbox{CdMnTe/CdMgTe}
quantum well (QW) buried at nanometers distances ($d$) below two
types of experimentally important magnetic Fe islands: with
rectangular \cite{Kossut1} and cylindrical \cite{Berciu1} shape. In
order to make localization effects sizeable we used diluted magnetic
\mbox{CdMnTe} semiconductor (DMS) QW instead of classical
semiconductor because of a giant Zeeman effect that exist in DMS
(for a review see ref.~\cite{Furdyna1}). Theoretically, DMS can be
described by a Zeeman term with a very large g-factor. Dietl
\cite{Dietl1} \emph{et al.} reported electron \mbox{g-factor}
$|g_e|$=500 in sub-Kelvin experiment, which assuming known ratio of
exchange constant in CdMnTe: $\beta / \alpha = 4$, gives hole
\mbox{g-factor} $|g_h|$=2000. These giant, but realistic values of
\mbox{g-factors} were used in our calculations.
\begin{figure}[h]\label{figBzBoth}
  \includegraphics[height=.27\textheight]{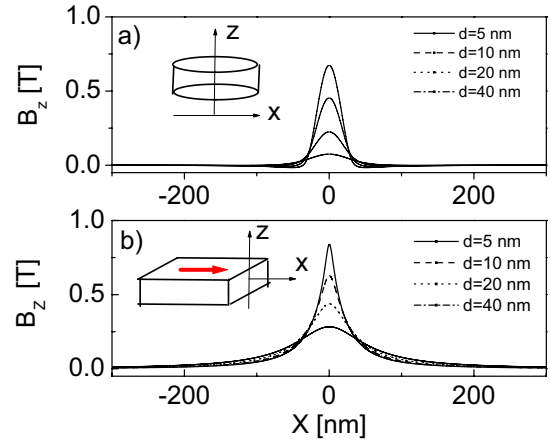}
 \caption{Upper panel: magnetic field $B_z$ at a distance $d$ below the center of magnetic
  micro-disk. Lower panel:  $B_z$ at distance $d$ below one of the two ends of the rectangular
  micro-magnet. Both magnets have the same thickness $D_z$=5 nm. With increasing $d$ maximum of $B_z$ is decreasing.}
\end{figure}

The magnetic field produced by the magnetic islands was calculated
by solving magneto-static equations with magnetization distribution
determined from micro-magnetic simulations in the case of
micro-disk, and by direct integration of Maxwell's magneto-static
equations in the case of the rectangular magnet, where homogenous
magnetization was assumed. Then, the energy spectrum of conduction
band as well as the spectrum of a valence band were calculated by
approximating the Schr\"{o}dinger differential eigen-equation with a
finite difference algebraic equation. In the case of
\mbox{$\mu$-disk}, valence band electrons were modeled by Luttinger
Hamiltonian. Inclusion of mixing between heavy and light holes is
important because of anisotropic spin-splitting \cite{Kuhn1} of
valence energy levels in DMS QW in the presence of an external
magnetic field with an arbitrary direction. Using Fermi's Golden
Rule, together with  previously calculated energies and wave
functions, we could calculate absorption coefficient $\alpha(E)$.
Each transition line was broadened with a Gaussian function in order
to simulate the realistic absorption data.

\begin{figure}[ht]\label{figAbsorDisk}
  \includegraphics[height=.25\textheight]{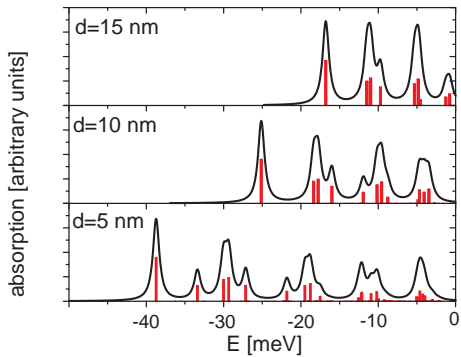}
  \caption{Absorption spectra (full lines) of a hybrid structures composed of a Fe
  magnetic island deposited on the surface of DMS QW structure at distance \mbox{$d$=15 nm},
  \mbox{$d$=10 nm}, and \mbox{$d$=5 nm} from QW. In our calculations each transition line (bar)
  was broaden by Gaussian function with line width of 1 meV.}
\end{figure}

We first consider a Fe micro-disk \cite{Berciu1} (diameter
\mbox{$R=1\,\mu$m} and thickness \mbox{$D_z$=50 nm},
\mbox{$\mu_0M_s=$2.2 T}) in the so called vortex state. In this
state the magnetization lies mainly in plane of the disk but in its
center it is forced out of plane. The diameter $R_c$ of the core,
where magnetization is pointing out of plane, extends only over 60
nm. It is important to mention that only $M_z$, the z-component of
the total magnetization $\vec{M}$, produces magnetic field $\vec{B}$
which couples to the quasi-particle's spin. The calculated profiles
of the $B_z$ component, that gives largest confinement effect, are
presented in Fig.~\ref{figBzBoth}~a. At a distance of $d$=10 nm
below the Fe magnet, the maximum magnetic field is $|B|_{max}$=0.46
T and its spatial extension is over a distance 80 nm. In
Fig.~\ref{figAbsorDisk} we present $\alpha(E)$ for three distances
$d$ between micro-magnetic disk and QW: \mbox{$d$=15 nm},
\mbox{$d$=10 nm}, and \mbox{$d$=5 nm}. Energy is measured relative
to the energy of the main absorption edge in QW without deposition
of magnetic island. Vertical bar height represent oscillator
strengths of the optical transitions. As expected, low-energy peaks
appear at lower energies with decreasing $d$ because the magnitude
of $\vec{B}$ increases. Separation between peaks is approximately
the same \mbox{(3 meV)} for all three distances $d$.

\begin{figure}[ht]\label{figAbsorRect}
  \includegraphics[height=.31\textheight]{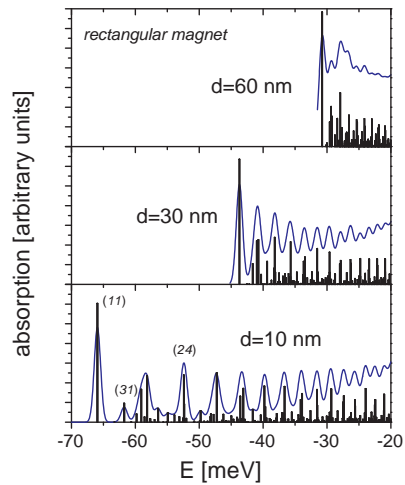}
  \caption{Absorption spectra (full lines) in the case of rectangular ferromagnetic island at
  three distances $d$ between island and QW. For \mbox{$d$=10 nm}, numbers ($nm$) indicate that given
  line corresponds to the transition from m-hole state to n-electron state.}
\end{figure}

Next, we analyze rectangular $\mu$-magnet \cite{Kossut1} (size:
\mbox{$D_x$=6 $\mu$m}, \mbox{$D_y$=2 $\mu$m}, and \mbox{$D_z$=0.15
$\mu$m}, $\mu_0M_s$=2.2 T) in a single domain state, with
magnetization pointing in x-direction. Please note that the
thickness of this magnet is 3 times larger than cylindrical
\mbox{$\mu$-disk} discussed  above. The corresponding field profile
is plotted in Fig.~\ref{figBzBoth}~b. The magnetic field is constant
along the $y$ edge of the magnet and therefore, the field produces
one dimensional confining potential. In Fig.~\ref{figAbsorRect} we
present the absorption spectra for three distances: \mbox{$d$=10
nm}, \mbox{$d$=30 nm}, and \mbox{$d$=60 nm.} In panel for
\mbox{$d$=10 nm}, the numbers ($nm$) indicate that corresponding
line is associated with transition from m-hole state to n-electron
state. The interesting property of field-induced confining potential
is that non-diagonal transitions ($n \neq m$) are very strong, and
can even be stronger than diagonal ones ($n=m$). For example,
transition ($24$) is stronger than transition ($33$), which is not
even visible in this scale.

Additionally, we have performed calculations for various magnet
thickness. We found that although the spatial extent of the field
increases for thicker micromagnet the effect of increasing amplitude
of the field is dominating and therefore the energetic distance
between ($11$) transition and main absorption peak of QW, as well as
distance between peaks in the spectrum increase with the increasing
micromagnet thickness. Therefore it is more promising to make hybrid
structures with relatively thick micromagnets.

This work was supported by the \mbox{NSF-NIRT} grant
\mbox{DMR02-01519}.



\begin{thebibliography}{5}
\expandafter\ifx\csname
natexlab\endcsname\relax\def\natexlab#1{#1}\fi
\providecommand{\enquote}[1]{``#1''} \expandafter\ifx\csname
url\endcsname\relax
  \def\url#1{\texttt{#1}}\fi
\expandafter\ifx\csname urlprefix\endcsname\relax\def\urlprefix{URL
}\fi

\bibitem[Kossut(2001)]{Kossut1}
Kossut, J., \emph{et al., App. Phys. Lett.}, \textbf{79}, 1789
(2001).

\bibitem[Berciu and Jank\'{o}(2003)]{Berciu1}
Berciu, M., and Jank\'{o}, B., \emph{Phys. Rev. Lett.}, \textbf{90},
246804
  (2003).

\bibitem[Furdyna and Kossut(1988)]{Furdyna1}
Furdyna, J.~K., and Kossut, J., \emph{Diluted Magnetic
Semiconductor},
  Academic, San Diego, 1988.

\bibitem[Dietl(1991)]{Dietl1}
Dietl, T., \emph{et al., Phys. Rev. B}, \textbf{43}, 3154 (1991).

\bibitem[Kuhn-Heinrich(1994)]{Kuhn1}
Kuhn-Heinrich, B., \emph{et al., Solid State Commun.}, \textbf{91},
413 (1994).

\end{thebibliography}
\end{document}